\documentclass{article} 

\usepackage[english]{babel} 

\usepackage[utf8]{inputenc}
\usepackage{graphicx}
\usepackage[usenames, dvipsnames]{color}
\usepackage{csquotes}
\usepackage{wrapfig}
\usepackage{lscape}
\usepackage{rotating}
\usepackage{epstopdf}
\usepackage{accents}
\usepackage{float}
\usepackage{bbm}
\usepackage[numbers]{natbib}
\newcommand{\ubar}[1]{\underaccent{\bar}{#1}}
\graphicspath{ {img/} }

\title{Evaluating Predictive Models of Student Success: Closing the Methodological Gap}

\author{Josh Gardner \\
        School of Information \\
       The University of Michigan\\
       jpgard@umich.edu
\and
Christopher Brooks \\
            School of Information \\
            The University of Michigan\\
       brooksch@umich.edu}

\begin{document}


\maketitle 



\begin{abstract}
    Model evaluation -- the process of making inferences about the performance of predictive models -- is a critical component of predictive modeling research in learning analytics. We survey the state of the practice with respect to model evaluation in learning analytics, which overwhelmingly uses only na{\"i}ve methods for model evaluation or statistical tests which are not appropriate for predictive model evaluation. We conduct a critical comparison of both null hypothesis significance testing (NHST) and a preferred Bayesian method for model evaluation. Finally, we apply three methods -- the na{\"i}ve average commonly used in learning analytics, NHST, and Bayesian -- to a predictive modeling experiment on a large set of MOOC data. We compare 96 different predictive models, including different feature sets, statistical modeling algorithms, and tuning hyperparameters for each, using this case study to demonstrate the different experimental conclusions these evaluation techniques provide.
\end{abstract}

\section{Introduction}\label{sec:introduction}

The past decade has seen an explosion in the use of data science methods in general, and predictive modeling in particular. Predictive modeling of student success has become a central task in learning analytics research. The increased use of digital learning tools such as Massive Open Online Courses (MOOCs) has made it easier than ever to store, replicate, transfer, and analyze learner data, and improvements in open-source statistical software for data science have generated an ``embarrassment of riches'' with seemingly limitless numbers of modeling techniques available for use. 

A critical scientific component of predictive modeling research is the process of \textit{model evaluation}, where inferences are drawn about the performance of a set of predictive models. In an optional step known as \textit{model selection}, a single preferred model is selected according to some objective function.

In this paper, we argue that there is a significant methodological gap in current practice in the learning analytics and educational data mining communities with respect to model evaluation and selection, and we present a survey of techniques to close this gap. We present data from a review of prior predictive modeling work in MOOCs to establish the ``state of the practice'' in the field in Section \ref{sec:state-practice}. In Section \ref{sec:model-evaluation}, we present a brief survey of methods for model evaluation, drawing from several fields outside of learning analytics in order to identify methods which are most effective, and which techniques currently used by the field are ineffective, for predictive model evaluation. In Section \ref{sec:case-study}, we conduct a predictive modeling experiment on a large sample of MOOCs, applying three techniques to demonstrate the different conclusions of three evaluation methods in practice. Our results also demonstrate specific empirical findings which are relevant to future modeling research in MOOCs, described in Section \ref{sec:results}. We provide conclusions with an eye to practice within the learning analytics field in Section \ref{sec:conclusion}.

\section{State of the Practice: Model Evaluation in Learning Analytics}\label{sec:state-practice}

The current work is concerned with experiments which construct multiple predictive models and attempt to compare their performance. In this section, we survey a large sample of 87 such experiments in MOOCs. Having passed the peer review process in several of the field's flagship journals and conferences (Computers and Human Behavior, Journal of Educational Data Mining, International Conference on Learning Analytics and Knowledge, The International Conference on Educational Data Mining, Learning at Scale, etc.), these results also reflect the consensus of reviewers -- and the field as a whole-- on techniques for predictive model evaluation in MOOCs.\footnote{The survey methodology is described in detail in \cite{Gardner2018-lp}; complete results of the literature review, including citation and categorization information for each study evaluated, is provided in the appendix of that work. Due to considerations of length we omit a thorough description here.}  This consensus is surprisingly strong, considering the breadth of approaches taken to other aspects of predictive modeling (e.g. feature extraction, statistical algorithms). The existing consensus is also on a set of techniques which are often statistically problematic, as we will discuss in Section \ref{sec:model-evaluation}. 

\subsection{Model Evaluation in MOOCs}\label{sec:mod-eval-moocs}

\subsubsection{Prediction Architecture}~\label{sec:lit-review-pred-architecture}

\begin{figure}
    \centering
    \includegraphics[width = \textwidth]{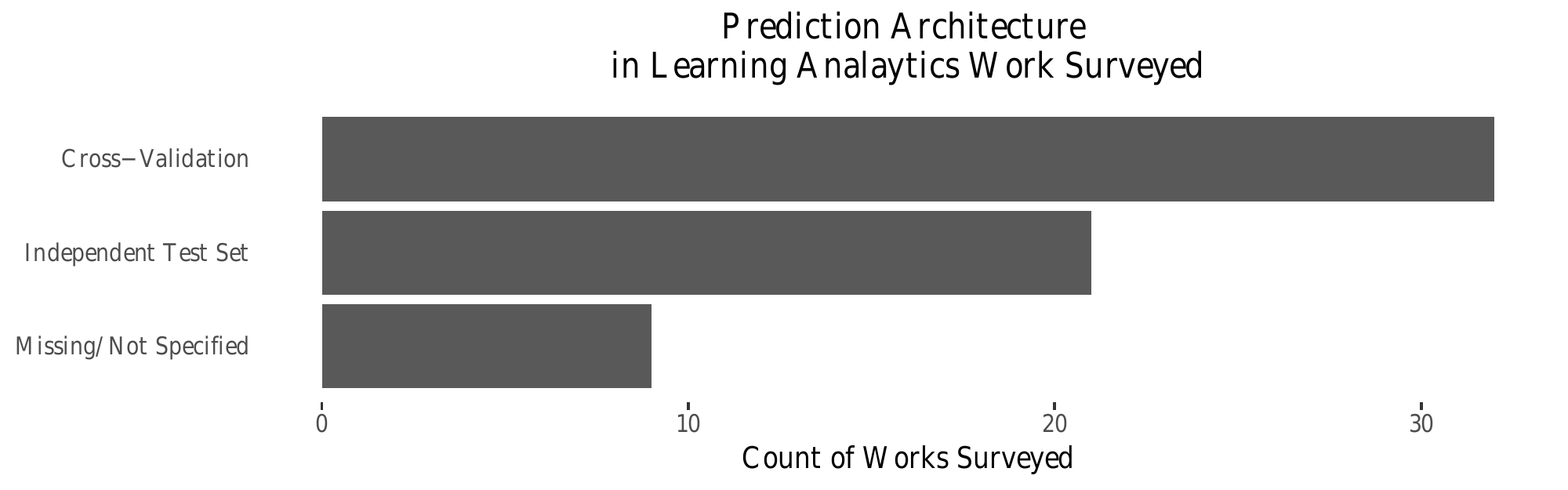}
    \label{fig:model-evaluation}
    \caption{Prediction architecture in works surveyed; this is the method by which estimates of model performance on unseen data are obtained (for further analysis, i.e., by applying statistical tests do the data).}
\end{figure}

In order to obtain the model and its predictions for evaluation, an experiment must choose a \textit{prediction architecture} -- the procedure by which training and testing datasets are partitioned and model predictions are obtained. The prediction architectures used in the works included in our  survey are shown in Figure \ref{fig:model-evaluation}.

These results demonstrate the prevalence of various forms of cross-validation in evaluating models (typically 2-, 5-, or 10-fold cross-validation; only two experiments surveyed use leave-one-out cross-validation). We found no cases where fold-level data was reported or evaluated (for example, to produce estimates of the variance of model performance across each fold). Model performance data was only evaluated and reported as an average for all works surveyed. The use of cross-validation is particularly relevant in light of concerns about statistical testing applied to cross-validated model performance data discussed in Section \ref{sec:model-evaluation}. In nine experiments (nearly 10\% of the experiments surveyed), the prediction architecture was not even reported, despite the presentation of predictive results in the work. 

\subsubsection{Model Evaluation}~\label{sec:lit-review-model-evaluation}

Model evaluation is the procedure by which observed differences in predictive performance are formally evaluated in order to draw inferences from the results of an experiment. We present data on model evaluation methods for any experiments where multiple predictive models were evaluated and compared, and where the authors reported inferences about which model was a ``better'' predictor of the outcome of interest, in Table \ref{tab:model-testing}. Note that strictly explanatory modeling experiments, where inspection of the model itself is the stated primary goal, are under the ``No model comparison or explanatory'' label. While our categorization cannot be perfect based only on the published descriptions of work, the results in Table \ref{tab:model-testing} make clear the lack of use of statistical methods for model evaluation in current learning analytics research -- a practice which we believe is scientifically problematic. Below, we will also demonstrate that the most prevalent methods for model evaluation shown in Table \ref{tab:model-testing} are known to be ineffective for the contexts in which they are being used (e.g., with experimental data from a cross-validation architecture; for large numbers of comparisons).

\begin{table}[]
\centering
\begin{tabular}{p{6.5cm} p{1cm}}
\hline
Student's $t$-test & 5 \\ 
Other NHST (Chi-Square, etc.) & 5 \\ 
No model comparison or explanatory & 24 \\
\textcolor{red}{Model comparison with no statistical test} & \textcolor{red}{51} \\ \hline
Total & 87 \\ \hline
\end{tabular}
\caption{Testing procedures for model evaluation in works surveyed.}
\label{tab:model-testing}
\end{table}

This consensus on techniques for predictive model evaluation in the field of learning analytics is as surprising as it is strong. Cross-validation with no statistical testing is overwhelmingly adopted in the learning analytics and educational data mining predictive modeling research, to the point where the results of this procedure are considered sufficiently acceptable by authorities in the field to be published widely, despite known inferential flaws with such a procedure (discussed in Section \ref{sec:model-evaluation}). 

Furthermore, the impact of these methods is compounded by large numbers of hypotheses being tested: Fewer than 20\% of the works surveyed compare one or two models; the remainder compared more than two models -- with the total number of comparisons unreported in an additional 20\% of cases. For example, \cite{Whitehill2017-tt} compares at least 1400 models with various architectures and window formulations; \cite{Taylor2014-hu} evaluates over 10,000 candidate models; neither correct or even acknowledge the large number of multiple comparisons in the context of their analysis. None of the works surveyed applied corrections for multiple testing, despite the fact that uncorrected multiple testing leads to elevated error rates, and existing methods were developed in the 1960s with ``between 2 and perhaps 20'' tests in mind and are not appropriate or effective for testing thousands of hypotheses \citep[][pp. 273]{Efron2016-to}.

\subsubsection{Hyperparameter Tuning}~\label{sec:lit-review-hyperparameter-tuning}

There is no broad consensus in practice regarding which feature extraction methods or algorithms should be used to construct models for student success prediction in MOOCs \cite{Gardner2018-lp}. However, almost all algorithms require selecting and, optionally, tuning hyperparameters which control elements of model fit and therefore the model parameters selected when the algorithm is applied to a specific dataset. These hyperparameters control convergence of parameter estimates, feature selection or regularization, and the model loss or ``cost'' function. Little attention has been paid to hyperparameter tuning in prior work, despite the fact that tuning each additional hyperparameter setting adds an additional hypothesis test or pairwise comparison to an experiment.

Our survey finds that in practice, hyperparameter tuning is frequently not described in published research: in 20 out of 87 (23\%) of the works surveyed, methods for hyperparameter tuning or selection were not reported or even mentioned, and in a remaining 9 (10\%), hyperparameter tuning was reported as being performed manually, with no reproducible procedure offered. In cases where hyperparameter tuning is not reported, either the hyperparameter tuning is (a) simply not performed, and default settings are used; (b) performed by the experimenter, but not reported, or (c) performed automatically by a machine learning software toolkit. As we will describe in detail below, (b) and (c) require model evaluation methods which are robust to multiple comparisons even \textit{without} knowing or reporting how many comparisons are conducted in an experiment. In (a), we might question whether an experiment sufficiently explored the performance of a given modeling approach; however, our case study in Section \ref{sec:case-study} showed little impact of hyperparameter tuning relative to the effects of different features or statistical algorithms, particularly when considering that tuning increases the number of models considered in a multiplicative fashion (because an experiment typically tests each algorithm/hyperparameter pairing with each set of features).

\subsection{Why is Model Evaluation Rare in Learning Analytics?}~\label{sec:why-model-eval-rare}

While it is difficult to determine with certainty, the previous analysis raises the question of why learning analytics has arrived at the status quo of inadequate model evaluation, several factors have likely contributed to the current state of the practice, including:

\textbf{Inadequate Tooling.} Software for model evaluation is rare, and is currently not included in (or even integrated with) many of the most common statistical software tools for machine learning research. Tracking the entire process of model exploration in an experiment therefore requires extensive manual effort or the use of additional specialized tools (e.g. \cite{Vartak2016-cu}). Furthermore, many machine learning toolkits ignore statistical model evaluation by design, and conduct intensive searches of the model and hyperparameter space without regard to statistical inference procedures (e.g. \texttt{AutoWEKA}, \texttt{SuperLearner} and \texttt{caret} in R, \texttt{auto-sklearn} in Python).

\textbf{Evolution of Data Science.} Many of the issues addressed in the current work were simply not concerns even a decade ago. Open-source data science methods, multicore computing, and statistical software have exploded in the time that learning analytics (and particularly MOOCs) have also come of age. As such, methods which addressed the research challenges of previous eras have become overwhelmed by these`computer age statistical inference'' problems, including the challenge of ``large scale hypothesis testing'' when the number of comparisons is massive \cite{Efron2016-to}.

\textbf{Lack of Training and Theory.} Many researchers who utilize predictive modeling simply were not trained in appropriate research methods for large-scale machine learning experiments; indeed, the theory guiding these methods is still the subject of active research \cite{Efron2016-to}.

\textbf{Lack of Incentive.} As the prior review demonstrates, there is currently little incentive to perform rigorous model evaluation in published work: reviewers do not expect it. Predictive modeling experiments can be published without the rigorous statistical evaluation that is expected of many other types of experimental work (e.g. randomized trials). Such evaluation may be perceived as even reducing the chances of publication by revealing ``insignificant'' results.

The above factors reflect the advancement of statistical and modeling capabilities which have not yet been matched by an advancement in methods or accepted practices.  We hope that, through works such as the current experiment, we can contribute to gradual progress on each; of course, future work is needed to address each of these more deeply.

\section{Predictive Model Evaluation and Selection}~\label{sec:model-evaluation}

Our focus in this section is to survey prior work on statistical model evaluation techniques which are (a) widely used in the field of learning analytics, or (b) particularly effective for tasks commonly faced in this field (e.g. evaluating multiple models across multiple datasets). We mention additional methods for model evaluation only where relevant; an exhaustive survey is beyond the scope of the current work. We conclude by presenting a series of criteria for an acceptable model evaluation procedure in Section \ref{sec:recommended-approach}, which lead us to prefer the Bayesian procedure for predictive model evaluation in learning analytics.

\subsection{Na{\"i}ve Average Method}\label{sec:sample-average}

The most common technique for comparing the performance of predictive models in MOOC research is what we term the ``na{\"i}ve average'' method. In this approach, models are evaluated by comparing their average performance (often, averaged across cross-validation folds) with no statistical test. As shown in Table \ref{tab:model-testing}, 51 out of the 87 studies (59\%) surveyed use the na{\"i}ve average method to draw inferences about a comparison between multiple predictive models (either implicitly, by presenting the predictive results from several models, or explicitly, by referring to models as ``more accurate'' or having the ``best performance''). Drawing inferences about which model may be ``best'' based on a simple sample average or the observed rankings is inappropriate for machine learned models for the same reasons it is inappropriate for drawing inferences from any other data: it provides no robustness against spurious results, no measure of confidence in the conclusion given the observed data, and little basis for comparison across studies. We call this method the ``na{\"i}ve average'' method because, by simply choosing the ``best'' average performance, this approach na{\"i}vely assumes that any differences observed must be (a) due to genuine differences in model performance (and not, for instance, random variation), and (b) that these differences must be both practically significant (large or important enough to be useful) and (in frequentist terms) statistically significant.

Statistical testing was developed to draw principled, reliable inductive inferences from data under uncertainty. This is particularly important when evaluating the complex performance data from predictive models of student success, which itself reflects underlying samples of student populations, randomized resamples of subpopulations (for instance, via cross-validation), and other stochastic procedures inherent to many modeling algorithms (such as random feature selection or parameter initialization methods). Failing to utilize any testing in the presence of intentional randomization makes the observation of spurious results more likely, and in the worst cases, could allow for exploitation of randomization to produce desired results (this behavior has been observed in other fields, such as reinforcement learning, where random seed ``optimization'' has been cited as a threat to reproducibility in the field \cite{Islam2017-gf}).

Furthermore, averaging itself may be particularly uninformative or misleading for model evaluation. The na{\"i}ve average method must assume \textit{commensurability across datasets} in order to justify the use of averaging. In the case of predictive models of student success, this assumption may not be justified: some courses might be easier or more difficult to predict on for a variety of reasons, including variability in student subpopulations, level of difficulty, quality of instruction, course durations and requirements, etc. Dem\v{s}ar notes that ``[i]f the results on different data sets are not comparable, their averages are meaningless'' \citep[][pp. 6]{Demsar2006-cx}. Averages are also susceptible to outliers, which particularly distort experimental results when relatively small populations are used (as is the case in our review, where nearly 50\% of studies evaluated only a single course). Furthermore, average cross-validated performance can generate ``optimistically'' biased model performance estimates \cite{Friedman2001-dh}, because models are repeatedly trained and tested on overlapping subsets of the same dataset, not a disjoint, truly unseen dataset as in a prediction architecture using an independent test set. 


\subsection{Null Hypothesis Significance Testing (NHST)}\label{sec:nhst}

In this section, we discuss model evaluation procedures based on null hypothesis significance testing (NHST). As Table \ref{tab:model-testing} demonstrates, this is the second most common model evaluation method in learning analytics.

A great deal of early work on model evaluation and selection in the field of machine learning focused on selecting from two models evaluated on a single dataset \citep[e.g.][]{Dietterich1998-vh}. While the field of learning analytics often addresses tasks which are more complex (multiple models and datasets), this early work highlighted several issues which any model evaluation procedure must address. These include having an acceptable error rate \cite{Bouckaert2003-rj, Dietterich1998-vh}, using methods which provide inferential replicability \cite{Bouckaert2004-xd, Bouckaert2004-oy}, and accurate estimation of statistical parameters in the presence of correlated data (such as the data generated by cross-validation) including the standard error \citep[e.g.][]{Nadeau2003-oq} and degrees of freedom \cite{Wang2016-pd}.

One two-model test is worth discussion here, due to its wide use in learning analytics research, as shown in Table \ref{tab:model-testing}: the Student's $t$-test. The $t$-test has been analyzed at length for predictive model evaluation, and many variants of the $t$-test have been roundly rejected as inappropriate and misleading for evaluating cross-validated model performance data, to which it is often applied in learning analytics. An important analysis was presented two decades ago in \cite{Dietterich1998-vh}, where a series of experiments demonstrated that several variants of the $t$-test (difference in proportions, paired $t$-test using repeated random subsampling, and $t$-test with 10-fold cross-validation) display an elevated Type I error rate and are therefore ineffective for model evaluation. Other empirical work has shown similar results with $t$-tests with a variety of other resampling and cross-validation architectures \citep[e.g.][]{Bouckaert2004-oy}. While various corrections have been proposed to this test \citep[e.g.][]{Nadeau2003-oq} and other alternatives exist \citep[e.g. the sorted runs scheme explored in][]{Bouckaert2003-rj}, they are rarely used in practice and were not used in any of the works surveyed, and are therefore not considered in detail here.

In practice, comparison of only two models on a single dataset account for only 10 (11\%) of the works surveyed, and it would be preferable to use model evaluation methods which can be easily extended to experiments with more than a single dataset. One method to adapt a $t$-test or other significance tests (e.g. Chi-square) to the multiple-comparisons case involves making an adjustment for multiple testing (which were not applied in any of the works surveyed which utilized these NHST procedures), such as that proposed in \cite{Benjamini1995-nu}. However, such procedures are not considered appropriate for cases where $k >> 20$ \cite{Efron2016-to}, and additional procedures for the multiple-model-multiple-dataset task exist which both address the multiple testing concerns and issues with parameter estimation mentioned above, as well as broader concerns with the suitability of NHST procedures such as the $t$-test.

The assumptions of traditional statistical tests used for data analysis, such as the paired $t$-test, are often strongly violated by predictive model performance data, which makes them particularly ineffective for evaluating hypotheses about predictive models. For example, the classical statistical procedure to determine whether there is a difference between several experimental subsamples is analysis of variance (ANOVA) \cite{Fisher1925-mv}, perhaps with a post-hoc test if groupwise differences were detected. This procedure is unfit for predictive model evaluation, however, because its assumptions of normality, sphericity, and independence (or non-correlation) are not guaranteed -- and frequently violated -- by the data from predictive modeling experiments \cite{Demsar2006-cx}. We do not discuss the fitness of ANOVA for model evaluation further because it was not used in any of the work surveyed, but refer the reader to \cite{Demsar2006-cx, Japkowicz2011-lw} for further discussion. 

Nonparametric procedures are often more appropriate when the assumptions of parametric procedures (such as the $t$-test) are likely to be violated. The Friedman test, in particular, has gained increasing adoption across the field of machine learning for model evaluation \cite{Demsar2006-cx, Japkowicz2011-lw}. The Friedman test is a nonparametric version of the ANOVA test, and compares the average rankings of the $k$ algorithms across each of $N$ datasets, calculating a test statistic measuring the probability of the observed rankings under the null hypothesis of all algorithms having equivalent performance (and therefore equal expected average rankings). The observed value of the Friedman statistic

\begin{equation}
    \chi_F^2 = \frac{12N}{k(k+1)} \left[ \sum_{j}R_j^2-\frac{k(k+1)^2}{4} \right]
    \label{eqn:friedman}
\end{equation}

where $R_j$ is the rank of the $j$th of $k$ algorithms on $N$ datasets and the statistic is distributed according to a chi-square distribution with $k-1$ degrees of freedom, is compared to a critical value for the given values of $N$ and $k$ \cite{Friedman1940-bd}. 

If the null hypothesis is rejected at the selected significance level, the post-hoc Nemenyi test is used to compare all classifiers to each other. The Nemenyi test is similar to a nonparametric version of the Tukey test for ANOVA, and uses a critical difference 

\begin{equation}
    CD = q_\alpha\sqrt{\frac{k(k+1)}{6N}}
    \label{eqn:CD}
\end{equation}

as a threshold to determine whether the performance between any two classifiers is significantly different, where the critical value $q_\alpha$ is based on the Studentized range statistic divided by $\sqrt{2}$.

\begin{figure}
    \centering
    \includegraphics[width = 4in]{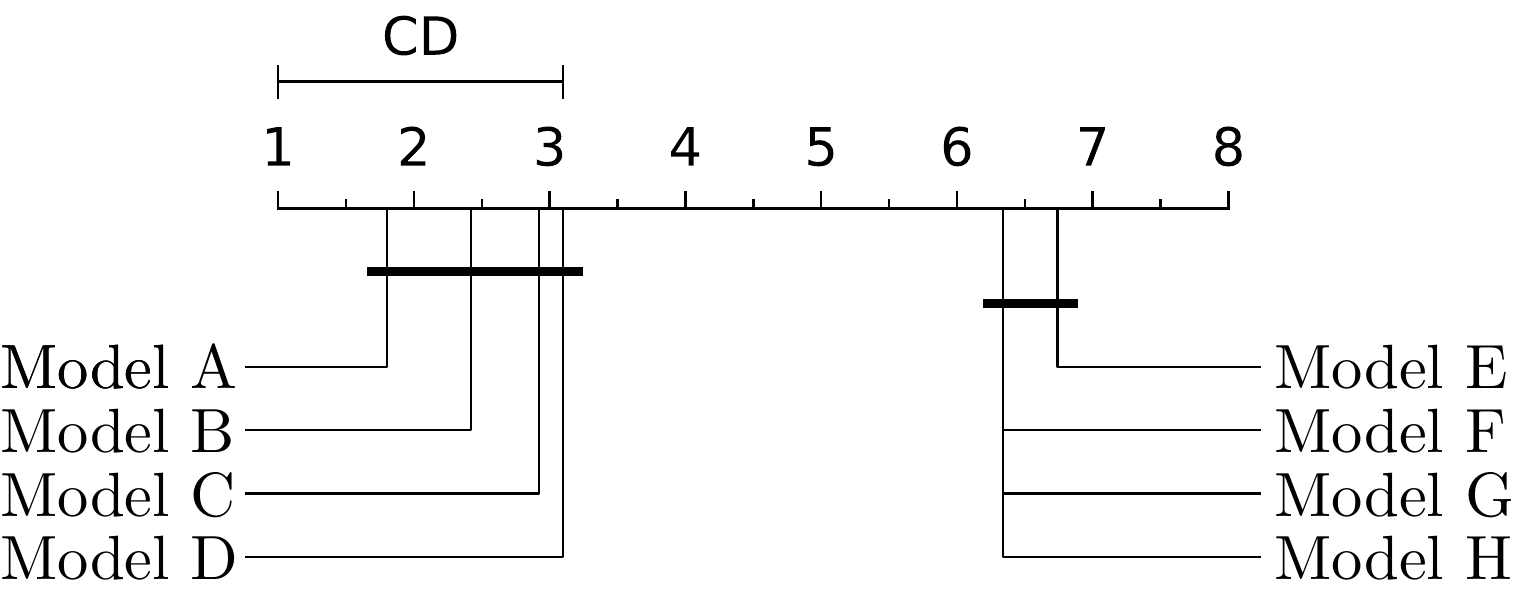}
    \caption{A Critical Difference (CD) diagram, which is used to visualize the results of the Friedman + Nemenyi testing procedure.}
    \label{fig:cd-example}
\end{figure}

The results of the Friedman and Nemenyi tests are often visualized using a Critical Difference (CD) diagram, originally proposed in \cite{Demsar2006-cx}. An example is shown in Figure \ref{fig:cd-example}. Models are plotted on a number line according to their average rank across all datasets, and bold CD lines are used to link models which are statistically indistinguishable at $\alpha$.

One advantage of this method is that because the Friedman test uses only the \textit{rankings} of the algorithms on each dataset, it does not require estimates of the variance of model performance -- recall that inaccurate estimates of variance were one of the primary confounding issues with $t$-tests applied to cross-validated model performance data. Instead, it only requires that the estimates of model performance and the measured rankings they produce are reliable and ``...that enough experiments were done on each data set and, preferably, that all the algorithms were evaluated using the same random samples'' \citep[][pp. 2]{Demsar2006-cx} and that the datasets, and therefore the rankings of the algorithms across each dataset, are independent. In contrast to many other statistical approaches to comparing model performance, such as ANOVA, the Friedman test makes no further assumptions about the sampling scheme.

This procedure also accounts for multiple comparisons. The number of models compared, $k$, is accounted for in both the Friedman statistic (Equation \ref{eqn:friedman}) and the post-hoc Nemenyi test (Equation \ref{eqn:CD}). The number of comparisons conducted in the course of a model evaluation experiment can grow quite large even with modest numbers of feature sets, algorithms, and hyperparameter settings (for example, in the case study in Section \ref{sec:case-study}, $k = 96$ and the number of pairwise comparisons is $\frac{96 \times 95}{2} = 4560$); adequate controls to moderate the error rate and inferential replicability are necessary for any test suited for model evaluation. For an example of the use of this method in another domain, see \cite{Madjarov2012-bo}.

\subsection{Bayesian Model Evaluation}~\label{sec:bayesian-estimation}

The application of Bayesian statistical methods to model evaluation has increased over the past two decades as scientific consensus around the concerns outlined above has grown, and as Bayesian modeling techniques and the computational infrastructure necessary to conduct them have become more widely accessible to researchers. There are several approaches to model evaluation which use Bayesian techniques, and we refer the reader to \cite{Benavoli2017-ff} for a review. In this work, we focus on an approach which Benavoli et al. refer to as \textit{Bayesian parameter estimation} or simply \textit{Bayesian analysis} \cite{Benavoli2017-ff}. We will refer to this approach as \textit{Bayesian model evaluation}, but occasionally use Benavoli et al.'s nomenclature when the meaning is clear.

Bayesian hierarchical models are used to address statistical applications which involve multiple parameters that can be regarded as related or connected. The hierarchical model encodes the dependence between these parameters such that certain aspects of the model depend on other parameters, which are referred to as \textit{hyperparameters}. The distributions of these hyperparameters are referred to as \textit{prior distributions} in this section to avoid confusion with the hyperparameters of machine learning models discussed in other sections. In a hierarchical model, the data are used to estimate the distribution of all parameters by either direct evaluation (which is rare) or by Markov Chain Monte Carlo (MCMC) sampling. As with any Bayesian model, a hierarchical model treats the parameters as random; we therefore estimate and explore their distribution instead of testing hypotheses about the ``true'' value of the parameter.\footnote{While a detailed introduction of Bayesian hierarchical modeling is beyond the scope of this paper, we refer the reader to \citep[][Ch. 5]{Gelman2014-wu} or \citep[][Ch. 9]{Kruschke2014-uq} for a thorough introduction.}

In the current work, we consider the Bayesian hierarchical correlated $t$-test of \cite{Corani2017-nx}. This test compares the results of multiple classifiers over multiple datasets when cross-validation is used. This test uses the following hierarchical model to account for the mean, variance, and correlation of the results of fold-level model performance data (an example of the data used for such a model is shown in Table \ref{tab:sample-data}):

\begin{equation}
    \mu_1, \ldots, \mu_k \sim t(\mu_0,\sigma_0,\nu)\label{eqn:bhctt-2}
\end{equation}

\begin{equation}
    \sigma_1, \ldots ,\sigma_k \sim Uniform(0, \bar{\sigma})\label{eqn:bhctt-3}
\end{equation}

\begin{equation}
    x_i \sim MVN(\mathbbm{1}\mu_i, \Sigma_i)\label{eqn:bhctt-1}
\end{equation}

Equation (\ref{eqn:bhctt-1}), where $x_i$ is the vector of differences between models, captures the correlation between cross-validation measures on the $i$th dataset by modeling these as draws from a multivariate normal distribution with mean $\mu_i$, and correlation $\rho$. $\mathbbm{1}$ is a vector of ones, and the covariance matrix $\Sigma$ has diagonal elements $\sigma_i^2$ and off-diagonal elements $\rho \sigma_i^2$ where $\rho = \frac{n_te}{n_tr}$, following Nadeau and Bengio \cite{Corani2017-nx, Nadeau2003-oq}. Equation (\ref{eqn:bhctt-3}) allows each dataset to have its own estimation uncertainty, standard deviation $\sigma_i$, drawn from a common uniform distribution where $\bar{\sigma} = 1000 \cdot \sum_{i=1}^q\frac{\hat{\sigma_i}}{k}$. We refer the reader to \cite{Benavoli2017-ff, Corani2017-nx} for further details.

Equation (\ref{eqn:bhctt-2}) models the differences between two classifiers on each dataset, and thus allows that some classifiers might perform better or worse on certain datasets, leading to variability in the difference $\mu_i$ on each dataset $i$. Equation (\ref{eqn:bhctt-2}) also models the fact that each observed mean difference on a single dataset, $\mu_i$, depends on the average difference of accuracy between the two classifiers on the population of data sets, $\mu_0$. This parameter, $\mu_0$, is typically the quantity of interest, and is modeled with a $t$-distribution with variance $\sigma_0^2$ and degrees of freedom $\nu$. The use of a $t$-distribution here makes the model more robust to outliers \cite{Kruschke2013-vv}, and slightly more conservative than its frequentist counterpart \cite{Corani2017-nx}.

Prior distributions for the model are given by:

\begin{equation}
    \sigma_0 \sim Uniform(0, \bar{s}_0),\label{eqn:bhctt-4}
\end{equation}

\begin{equation}
    \mu_0 \sim Uniform(-1, 1),\label{eqn:bhctt-5}
\end{equation}

\begin{equation}
    \nu \sim Gamma(\alpha, \beta)\label{eqn:bhctt-6}
\end{equation}

\begin{equation}
    \alpha \sim Uniform(\ubar{\alpha} = 0.5, \bar{\alpha} = 5)\label{eqn:bhctt-7}
\end{equation}

\begin{equation}
    \beta \sim Uniform(\ubar{\beta} = 0.05, \bar{\beta} = 0. 15)\label{eqn:bhctt-8}
\end{equation}

A detailed discussion of these priors is beyond the scope of this work, but we note that these are considered appropriate for most measures of predictive model performance when used with this procedure, including AUC, the measure used in the experiment below. The reader is referred to \cite{Corani2017-nx} for a detailed discussion.

This Bayesian hierarchical model is used to make inferences about average differences in model performance for each pair of candidate models $X$ and $Y$. From the fitted model, MCMC is used to generate samples of $\theta = (P(X>Y), P(ROPE), P(X<Y))$, where $\theta_i$ represents the posterior probability that model X is better, the models are equivalent, and model Y is better, respectively. These samples represent the hypothetical differences in performance on a future unseen dataset \cite{Corani2017-nx}; generating $N = 50,000$ samples on a typical laptop computer takes only a few seconds, and conducting this comparison for all 4560 pairwise comparisons in the experiment below takes less than 10 minutes using the \texttt{BayesianTestsML} Python library.\footnote{https://github.com/BayesianTestsML} 

\begin{figure}
    \centering
    \includegraphics[width = 3.5in]{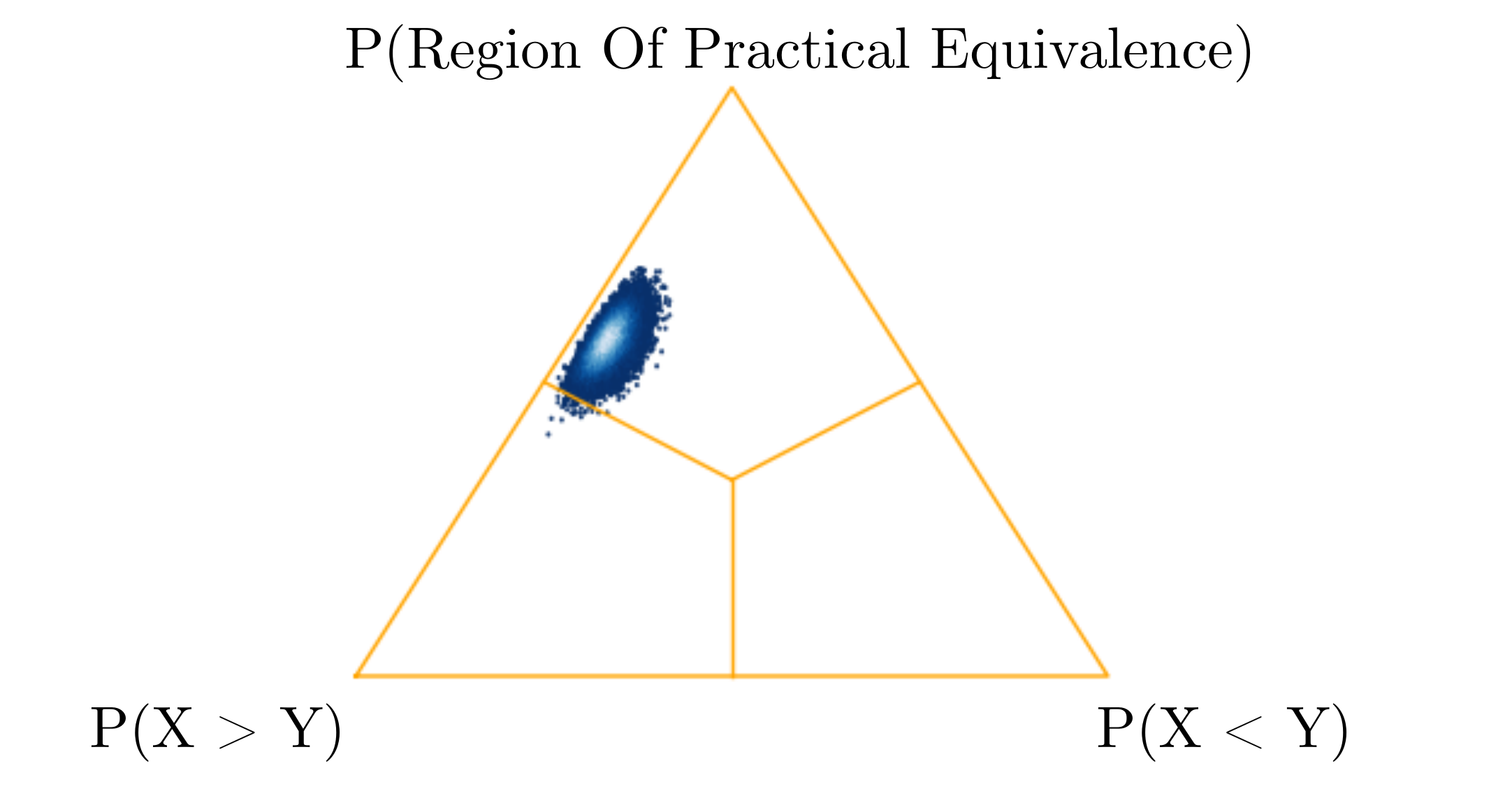}
    \caption{A Bayesian posterior plot resulting from a Bayesian hierarchical correlated $t$-test, which visualizes the results of Markov-Chain Monte Carlo (MCMC) sampling for the comparison of two models $X$ and $Y$. The estimated probability of each outcome is the proportion of samples that fall in each section of the plot.}
    \label{fig:posterior-plot-example}
\end{figure}

The MCMC samples are used to estimate $\theta$ by simply counting the proportion of samples for which $\theta_i$ has the highest posterior probability. The results of this sampling can be visualized by projecting the ($\theta_{X>Y}, \theta_{ROPE}, \theta_{X<Y}$) triplets onto barycentric coordinates to produce a posterior plot, shown in Figure \ref{fig:posterior-plot-example}. Inspection of these plots can be useful for small comparisons, but for experiments with large models spaces, inspecting all $\frac{k(k-1)}{2}$ pairwise plots is impractical.

This method is able to account for the different uncertainty which characterizes each dataset by estimating unique parameters for each dataset, and, because the hierarchical model applies shrinkage to the $\mu_i$ values when estimating them jointly, it estimates them more accurately than previous approaches which use maximum likelihood estimation \cite{Corani2017-nx}.

\subsection{The Case for Bayesian Model Evaluation}\label{sec:recommended-approach}

In the survey presented above and in other work \cite{Gardner2018-lp}, we have described the common practices of predictive modeling experiments in learning analytics. These include (a) a massive space of potential models due to many data sources, feature types, and algorithms used; (b) relatively small collections of datasets, for example, even the largest prior MOOC studies of which we are aware evaluate around 40 MOOCs (i.e., \cite{Whitehill2017-tt, Evans2016-gj}) and (c) large individual datasets, which make repeated model-fitting undesirable, if not intractable. 

These, along with more general scientific considerations common to model evaluation in any domain, give rise to a series of desiderata which any ideal model evaluation procedure must meet. We introduce the criteria individually below, and describe how each the model evaluation methods discussed above do or do not satisfy these criteria.

\textbf{Computationally tractable}: Statistical testing should require as little computational overhead as possible. It should scale well with the number of datasets $N$, the number of models $k$, and the dimensionality of the data. This consideration does not particularly count in favor of any of the schemes considered, but also does not rule any out. All are computationally tractible with consumer-level computing hardware. The MCMC sampling method used by the Bayesian method incurs a higher computational overhead than  NHST, but for the modest $N$ used in most MOOC research (at most $N = 44$ in the largest known prior analysis \cite{Evans2016-gj}), the computational cost is negligible. For equally effective schemes, we might decide to choose the one which can provide the most efficient estimates.

\textbf{Impose minimal assumptions on the data}: An ideal method will make few assumptions about the underlying data, such as normality, symmetry, commensurability across datasets, sphericity, etc. This alone excludes many common parametric NHST methods, such the $t$-test, ANOVA, or the na{\"i}ve average method, which make strong assumptions which are not met by model performance data in practice \cite{Demsar2006-cx}. While this excludes most parametric NHST methods, it does not provide a clear reason to prefer nonparametric NHST over Bayesianism.

\textbf{Account for cross-validation}: As discussed in section \ref{sec:nhst}, the use of cross-validation requires correction for the overlap between the training data used in each fold. This is particularly important because cross-validation is widely used in learning analytics. As discussed in Section \ref{sec:nhst}, the $t$-test in particular shows highly elevated inflated Type I error rates and low inferential replicability with cross-validated model performance data. To avoid this, the method in Section \ref{sec:nhst} ranks models over each dataset, ignoring important information in the fold-level performance data (such as the variability of model performance across folds). The Bayesian procedure, in contrast, utilizes the fold-level data directly to compute estimates of the variability of each pairwise model comparison, as shown in Equation \ref{eqn:bhctt-1}.

\textbf{Robust to multiple comparisons} As discussed in Section \ref{sec:lit-review-model-evaluation}, more than 80\% of the works surveyed compared more than two models. The practice of comparing many models is also often useful, allowing for the exploration of a diverse model space, providing many reference points, and allowing for benchmarking relative to other work. Model evaluation should therefore allow for many comparisons with minimal effect on inferential error rates. When existing NHST procedures are adjusted for multiple comparisons, these adjustments are often impractically conservative with large $k$ and small to moderate $N$ to avoid Type I error, as our case study demonstrates in Section \ref{sec:results} (see also \cite{Efron2016-to}). In contrast, the Bayesian approach does not ``accept'' or ``reject'' hypotheses and is generally unconcerned with Type I errors as it only estimates posterior probabilities. A Bayesian hierarchical model can directly account for the uncertainty from multiple comparisons, and applies shrinkage to estimators to account for this uncertainty \cite{Gelman2012-ys}.

\textbf{Test an informative $H_0$}: In the case of the NHST in Section \ref{sec:nhst}, we are testing a series of hypotheses of pairwise equivalence between the candidate models. That is, we are evaluating $H_0$: the performance of models $X$ and $Y$ are exactly equivalent (while the na{\"i}ve average does not test an $H_0$, here we could think of it as equivalent to an NHST that simply always rejects).  However, as noted previously, an $H_0$ of exactly equivalent performance is almost always false \cite{Benavoli2017-ff, Demsar2008-fy}. This $H_0$ has been called ``the \textit{nil} hypothesis'' after the probability that this hypothesis is true \citep[][pp. 1000]{Jensen1997-cn}. Particularly in the case of machine learning models, it is unlikely that any two algorithms have \textit{exactly} equivalent performance. Tests of this hypothesis, even when properly interpreted, are not indicative of any likely true state of the world, and are potentially misleading.

\textbf{Provide direct evidence about $H_0$}: When conducting model evaluation, we are interested in directly drawing inferences about some $H_0$. NHST cannot prove the null hypothesis or provide direct evidence of it \cite{Demsar2008-fy, Wasserstein2016-gm}, although they are commonly misinterpreted as doing so both in the broader scientific literature \cite{Cohen1994-cs, Demsar2008-fy, Cohen1994-cs} and in several of the MOOC experiments surveyed here. The Bayesian procedure, in contrast, allows for direct inference about the probability of models having similar or equivalent performance, using the estimated posterior distribution of the difference in performance between two models. 

\textbf{Separate magnitude, uncertainty, and sample size during inference}: The effects of the magnitude of observed differences, the uncertainty of estimates, and the number of data points should be separable in model evaluation. However, both the na{\"i}ve average and NHST obscure the influence of each of these factors. The na{\"i}ve average does so by simply ignoring all three factors. An NHST only provides a $p$-value which reflects a mix of effect size and uncertainty \cite{McShane2017-rs}, even when the observed effects might be too small to be considered practically significant, or might be associated with a high level of uncertainty. Even if the procedure itself might separate these factors, the reported $p$-value does not allow the reader to differentiate between the magnitude of the effect, uncertainty, and sample size \cite{Benavoli2017-ff, Wasserstein2016-gm}. This leads to the further complication that detecting a ``significant'' difference under a null hypothesis of equivalence only requires collecting enough data (or conducting enough runs of cross-validation), under which conditions $H_0$ can always be rejected \cite{Cohen1994-cs, Demsar2008-fy, Dietterich1998-vh, McShane2017-rs, Wasserstein2016-gm, Witten2016-mm}. By using a region of practical equivalence (ROPE), the Bayesian method separately estimates the magnitude and variability of each estimated effect (here, differences in AUC between a pair of models).

Together, these criteria collectively mount a strong imperative in favor of Bayesian model evaluation.

\section{Case Study: Evaluating MOOC Dropout Models}\label{sec:case-study}

In this section, we present and evaluate the results of an experiment which constructs and compares several dropout models across a large sample of MOOCs using the three model evaluation methods discussed (na{\"i}ve average; NHST; Bayesian). Our goal in this section is twofold. First, such a side-by-side comparison stands not only to illuminate procedural and inferential differences between these approaches, but also demonstrate how they can produce different conclusions from the same underlying data in practice when evaluating many candidate models. Second, we hope to demonstrate interesting and useful results in this experiment: the data, feature extraction, and statistical modeling methods used here are common in applied predictive modeling research in learning analytics, but have not been collectively compared in previous work. For additional examples of case studies utilizing this method for classifier selection across multiple datasets, see \cite{Benavoli2017-ff, Corani2017-nx}.

\subsection{Data}

The data used in this experiment are a large and diverse sample of $N = 48$ sessions of MOOCs offered by the University of Michigan on Coursera. These courses are from several diverse domains, including science and technology, finance, healthcare, politics, and literature. A summary of the data for the courses used is shown in Table \ref{tab:course-summary}. The data are used to predict a binary dropout label indicating whether a user showed any activity in the final week of the course.

\begin{table}[!t]
\centering
\begin{tabular}{ll} \hline 
Metric & Value \\ \hline 
Number of Sessions & 48 \\
Number of Unique Courses & 17 \\
Total Number of Active Students & 117,028 \\
Total Number of Interactions & 2,479,900 \\
Average Number of Active Students  & 2490 (2391) \\
Average Length  in weeks & 9.8 (2.6) \\
Average Number of Unique Forum Posters  & 507 (447) \\
Average Number of Course Assignments  & 1.2 (0.9) \\
Average Number of Quizzes  & 18 (16.8) \\
Average Number of Human-Graded Quizzes & 1.6 (1.5) \\ \hline 
\end{tabular}
\caption{Summary statistics for courses used in case study. Standard deviations in parentheses.}~\label{tab:course-summary}
\end{table}

This dataset represents one of the largest MOOC dropout prediction studies to date in terms of number of courses evaluated.\footnote{The largest-scale work on predictive modeling in MOOCs to date are \cite{Li2016-ze, Liang2016-us}, which each build predictive models on 39 XuetangX MOOCs; and \cite{Whitehill2017-tt}, which builds predictive models on 40 HarvardX MOOCs.} This means the current experiment represents a reasonable upper bound for the number of datasets, $N$, that a MOOC predictive modeling experiment might utilize (which serves as a limiting factor in the NHST procedure, as our results demonstrate). 

\subsection{Experimental Setup}~\label{sec:experimental-setup}

\begin{figure}
    \centering
    \includegraphics[width=5in]{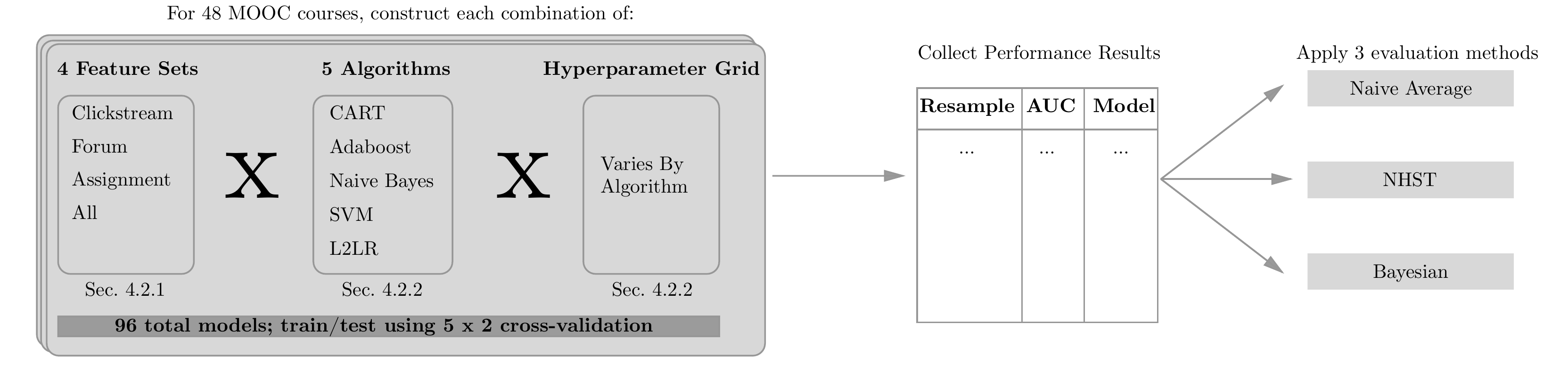}
    \caption{Experiment overview.}
    \label{fig:experiment-flow}
\end{figure}

The experimental setup, shown in Figure \ref{fig:experiment-flow}, is as follows: For each course session, we extract all four sets of features (clickstream, forum, assignments, all). We train each of the 96 candidate models on each feature set, using $5 \: \times \: 2$-fold cross-validation, using the same random cross-validation folds for every model, and recording the Area Under the Receiver Operating Characteristic curve (AUC) as the evaluation metric for each model. We retain the fold-level results (the model performance on the held-out model fold), which provides 10 estimates of model performance for every model on each session of the course. A sample of this data is shown in Table \ref{tab:sample-data}). We then apply the three methods for model evaluation to this data.

\subsubsection{Feature Extraction/Data Source} 

This process of transforming raw data into structured information suitable for use as input to a supervised learning algorithm is considered one of the most important and difficult tasks in predictive modeling in MOOCs \cite{Li2016-ze, Nagrecha2017-dn, Robinson2016-yr}. This experiment is designed to provide insight into which of these data sources within the Coursera platform \cite{Coursera2013-jh} might be most useful for MOOC dropout prediction. We evaluate four sets of features, each extracted from a single data source, with all individual features based on prior work:

\begin{itemize}
    \item \textbf{Clickstream}: Counting-based features representing the number of accesses to various course pages, number of forum views, and number of video views. These features are common in activity-based dropout modeling \citep[e.g.][]{Kloft2014-kb, Xing2016-le}. This is the simplest and smallest feature set.
    \item \textbf{Forum Posts}: Natural language processing-based metrics which measure sentiment, text complexity, and posting activity gathered from \cite{Crossley2016-ij, Robinson2016-yr, Wen2014-pv}. 
    \item \textbf{Assignments}: Academic performance metrics derived from students' quizzes, peer-graded assignments, in-video quizzes, and exams, including both simple features (e.g. average grade) as well as more complex features (e.g. number of submissions relative to the highest number of submissions by any student that week) \cite{Bote-Lorenzo2017-yh, Kotsiantis2010-hs, Veeramachaneni2014-ug}. Where courses used no assignments, models using this method defaulted to majority-class prediction.
    \item \textbf{All:} Union of the three features sets above.
\end{itemize}

All features and their definitions are shown in Table \ref{tab:feature-table}.

\begin{table}[!t] 
\centering
    \hspace*{-4.5cm}
    \begin{tabular}{p{6in}}
     \hline
    \multicolumn{1}{c}{\textbf{Clickstream}} \\
    \hline
    \textbf{Forum Views:} Number of pageviews of forum pages. \\
    \textbf{Active Days:} Number of days for which user registered any clickstream activity (maximum of 7). \\
    \textbf{Quiz Views:} Number of pageviews of quiz attempt pages, as measured by clickstream features. \\
     \textbf{Exam Views:} Number of pageviews of exam-type quiz pages, as measured by clickstream features. \\
     \textbf{Human-Graded Quiz Pageview:} Number of pageviews of human-graded quiz pages, as measured by clickstream features. \\ 
     \hline
    \multicolumn{1}{c}{\textbf{Assignments}} \\
    \hline
    \textbf{Pre-Submission Lead Time:} Time between a quiz submission and deadline for all submissions; discretized buckets for $t \geq 7$ days, $3 \leq t < 7$, $1 \leq t < 3$, $0 \leq t < 1$, and late. \\
     \textbf{Total Raw Points:} Sum of total raw points earned on quizzes. \\
     \textbf{Average Raw Score*:} Average raw score on all assignments. \\
     \textbf{Raw Points Per Submission:} Total raw points divided by total submissions. \\
     \textbf{Total quiz submissions:} Total count of quiz submissions. \\
     \textbf{Percent of allowed submissions:} Total count of quiz submissions as a percent of the maximum allowed submissions. \\
     \textbf{Percent of max student submissions:} A student total number of quiz submissions as a percent of the maximum number of submissions made by any student in the course. \\
     \textbf{Correct submissions percent*:} Percentage of the total submissions that were correct. \\
     \textbf{Change in weekly average*:} Difference between current week average and previous week average quiz grade. \\ 
      \hline
    \multicolumn{1}{c}{\textbf{Forum}} \\
    \hline
    \textbf{Number of Posts:} Total number of posts. \\
    \textbf{Number of Replies:} Number of posts by user which were replies to other users (i.e., not to themselves, and not first post in thread). \\
     \textbf{Average Post Sentiment:} Average net sentiment of posts (positive - negative) \cite{Hutto2014-po}. \\
     \textbf{Average Post Length:} Average length of posts, in characters. \\
     \textbf{Positive Posts:} Number of posts with net sentiment $ \geq 1$ standard deviation above thread average. \\
     \textbf{Negative Posts:} Number of posts with net sentiment $ \leq -1$ standard deviation below thread average. \\
     \textbf{Neutral Posts:} Number of posts with net sentiment within 1 standard deviation of thread average. \\
     \textbf{Sentiment Relative to Thread:} Average of (post sentiment - avg sentiment for thread). \\
     \textbf{Threads Started:} Total number of threads initiated by student. \\
     \textbf{Unique Words/Bigrams:} Count of unique words/bigrams used across all posts. \\
     \textbf{Flesch Reading Ease:} Flesch Reading Ease score, discretized into separate features in increments of 10 from 0 to 100 \cite{Kincaid1975-wz}. \\
     \textbf{Flesch-Kincaid Grade Level:} Flesch-Kincaid grade level, discretized into separate features in increments of 1 from 0 to 20 \cite{Kincaid1975-wz}. \\  
     \textbf{Net Votes Received:} Total net upvotes users' posts received (positive - negative). \\ \hline
    \end{tabular}\hspace*{-4cm}
    \caption{Feature name and definition by category. Each feature is calculated at the student-week level, resulting in $p \cdot n$ features at week $n$. Features marked with a (*) were calculated by quiz type (homework, quiz, and video), resulting in three different features, one per quiz type, using that definition.}
    \label{tab:feature-table}
\end{table}

\subsubsection{Algorithms and Hyperparameters} 

We consider the following models in our experiment: (1) classical decision trees (CART) \cite{Breiman1984-or}, (2) L2 (or ``ridge'') regularized logistic regression (L2LR); (3) gradient boosted tree (Adaboost) \cite{Culp2006-dd}, used as a a stand-in for the widely used \cite{Gardner2018-lp} random forest method\footnote{The random forest method did not allow us to test consistent values of the \texttt{mtry} parameter, number of variables to consider at each split, because this value depends on the number of variables in a dataset, and our experiment requires testing feature sets with different numbers of dimensions.}; (4) support vector machine (SVM) with linear kernel; (5) na{\"i}ve Bayes (NB). These represent five of the most commonly used modeling algorithm in predictive models of student success in MOOCs \cite{Gardner2018-lp}. A summary of the models considered, and any special preprocessing, is shown in Table \ref{tab:algorithms-hyperparams}. In addition to constituting a representative sample of the models most often used for dropout modeling tasks, these algorithms represent a broad spectrum of model types, including relatively simple, high-bias parametric models and complex, flexible, nonparametric models.

\begin{table}[]
\centering
\hspace*{-2cm}\begin{tabular}{llll} \hline 
\textbf{Algorithm} & \textbf{Hyperparameters Tuned} & \textbf{Total Models} \\ \hline 
Classification Tree & Cost-complexity parameter ($cp$) & 4  \\
Logistic Regression & L2 (ridge) penalty term ($\lambda$) & 5 \\
Adaboost & \begin{tabular}[c]{@{}l@{}}Boosting Algorithm (``Real Adaboost'', M1)\\ Number of iterations\end{tabular} & 6 & \\
SVM  \textasteriskcentered, \textdagger & Cost ($\gamma$) & 5 \\
Na{\"i}ve Bayes \textasteriskcentered  & \begin{tabular}[c]{@{}l@{}}Laplacian smoothing ($fL$) \\ Kernel\end{tabular} & 4\\ 
\textbf{Total} & & 24 \\
\hline 
 
\end{tabular}
\caption{Algorithms used and hyperparameters tuned for each. Each (algorithm, hyperparameter) set was used with each of the four feature types, yielding a total of $24 \times 4 = 96$ models. Preprocessing codes: \textasteriskcentered indicates zero-variance predictors, if any, were dropped for model training (as a requirement of model-fitting algorithms); \textdagger indicates predictors were centered and scaled.}~\label{tab:algorithms-hyperparams}
\end{table}

\begin{table}[!t]
\centering
\hspace*{-4cm}\begin{tabular}{lllll} \hline 
\textbf{AUC}         & \textbf{Resample}   & \textbf{Course}           & \textbf{Session}  & \textbf{Model}                          \\ \hline 
0.946 & Fold1.Rep1 & Digital Democracy & 1 & CART($cp = 0.01$, Features = All)\\ 
0.937 & Fold1.Rep1 & Digital Democracy & 1 & CART($cp = 0.1$, Features = All)\\ 
0.500 & Fold1.Rep1 & Digital Democracy & 1 & CART($cp = 1$, Features = All)\\
0.941 & Fold1.Rep2 & Digital Democracy & 1 & CART($cp = 0.001$, Features = All) \\
\ldots & \ldots & \ldots & \ldots & \ldots \\
0.748 & Fold1.Rep1 & Digital Democracy & 1 & L2LR ($\lambda = 0.1$, Features = Clickstream)        \\
0.742 & Fold1.Rep1 & Digital Democracy & 1 & L2LR ($\lambda = 0.01$, Features = Clickstream)       \\
0.743 & Fold1.Rep1 & Digital Democracy & 1 & L2LR ($\lambda = 0.001$, Features = Clickstream)      \\
0.701 & Fold2.Rep1 & Digital Democracy & 1 & L2LR ($\lambda = 1$, Features = Clickstream)          \\
\ldots & \ldots & \ldots & \ldots & \ldots \\
0.523 & Fold1.Rep1 & Digital Democracy & 1 & SVM ($\gamma = 10$, Features = Forum) \\
0.529 & Fold1.Rep1 & Digital Democracy & 1 & SVM ($\gamma = 1$, Features = Forum) \\
0.517 & Fold1.Rep1 & Digital Democracy & 1 & SVM ($\gamma = 0.1$, Features = Forum) \\
0.507 & Fold2.Rep1 & Digital Democracy & 1 & SVM ($\gamma = 10$, Features = Forum) \\
0.508 & Fold2.Rep1 & Digital Democracy & 1 & SVM ($\gamma = 1$, Features = Forum) \\
\hline 
\end{tabular}\hspace*{-4cm}
\caption{A sample of the experimental data generated. Each row represents a specific feature/algorithm/hyperparameter model on a specific cross-validation fold; all models were evaluated on identical folds.}~\label{tab:sample-data}
\end{table}

\section{Results Analysis}~\label{sec:results}

With four feature sets, 24 candidate models representing all algorithm/hyperparameter settings, and 10 iterations of cross-validation on 48 course sessions, this resulted in a total of $4 \times 24 \times 10 \times 48 = 46,080$ total observations of model performance.  We apply each method (na{\"i}ve average, NHST, and Bayesian) in order to identify the family of ``best'' models, which we define as:

\begin{quote}
    \textbf{Family of best models:} a set $\mathcal{F}$ of $m$ models, selected from a pool of $N$ candidate models, which have the best predicted performance on an unseen dataset. All models $f \in \mathcal{F}$ have equivalent (or practically equivalent) generalization performance to all others in $\mathcal{F}$, but better performance than any model not in $\mathcal{F}$. Models $f \in \mathcal{F}$ are sorted in descending order of performance, such that $\mathcal{F} =  \{f_1 > f_2 > \ldots > f_m\}$.
\end{quote}

All three model evaluation methods provide identical recommendations about the highest-performing model in this experiment (e.g. $f_{1,naive} = f_{1,NHST} = f_{1,Bayes}$): while differences are possible between the Bayesian method, which uses fold-level data, and the other two methods, which use average performance and rankings, the model performance was consistent enough that such differences in $\mathcal{F}$ were not observed here. However, these methods vary in terms of (a) the statistical inferences they support; (b) the size of $\mathcal{F}$, and (c) their ability to discriminate between pairwise comparisons across entire space of models, even those not within $\mathcal{F}$ (we explore this because experiments often seek to evaluate results besides the composition of $\mathcal{F}$). We demonstrate (i) the NHST method's limitations with the modest number of comparisons being performed on this dataset, which results in an inability to discern differences between more than 50\% of the models considered and a family of nearly 20 ``best'' models; (ii) that the Bayesian method generates the most robust and reliable inferences about model performance; and (iii) that the Bayesian method supports useful inferences about the relative performance of many features and algorithms for dropout prediction in MOOCs.

\subsection{Na{\"i}ve Average Method}

Recall in the na{\"i}ve average model evaluation approach, the experimental results are simply averaged by model, sorted according to performance, and the model with the best average performance is selected (assuming no ties).

Applying na{\"i}ve average method, we select $\mathcal{F}_{naive}$ as the single model with the highest average performance. This is the decision tree (CART) algorithm, with all features and a cost-complexity parameter of 0.001. The ``family'' of best models is thus a family of one (this corresponds to the top row of Table \ref{tab:frequentist-best-mods}). With the na{\"i}ve average method, every pairwise difference is considered significant, regardless of the magnitude of any observed difference, the total number of models tested, or the number of observations collected in the experiment: the observed differences are assumed to be accurate, not spurious, and large enough to be practically significant.

These results make clear how problematic the na{\"i}ve average method is. The difference in average AUC between $f_{1,naive}$ and the next-highest performing algorithm is less than 0.003 (see Table \ref{tab:bayesian-best-mods}). This difference is indeed small enough to be spurious, practically useless, or both. With $\frac{96 \times 95}{2}$ pairwise comparisons, we might expect to observe some differences in model performance due to randomness alone. Thinking in terms of Tukey's card analogy, the question of how many ``deals'' we made, and therefore how surprised we should be by the hand we have been dealt, is not considered by the na{\"i}ve average method.

Furthermore, the na{\"i}ve average method provides no principled estimate of the confidence or significance of our results, which provides no basis for comparison for future work which attempts to replicate these findings on new data: is our confidence low, in which case a different result would be surprising? Or is our confidence quite high, in which case we would expect similarly strong results in replications? The na{\"i}ve average method provides no answer.

Finally, because $\mathcal{F}_{naive}$ contains only one model, the na{\"i}ve average method does not allow us to easily introduce other considerations, such as model interpretability or training time, into our decision despite the fact that many models seem to have performance reasonably close to the highest-performing model. If we had a set of practically equivalent ``best'' models to choose from (if $\mathcal{F}$ contained more than a single element), we might select a final preferred model based on these other considerations. However, under the na{\"i}ve average method, we only ever have a single model in $\mathcal{F}_{naive}$, leaving no principled method for model selection even when the observed differences with other models are arbitrarily small.

\subsection{NHST Method: Frequentist Nemenyi Test}~\label{sec:nhst-case-study}

\begin{figure}[!t]
            \centering
            \includegraphics[width = \textwidth]{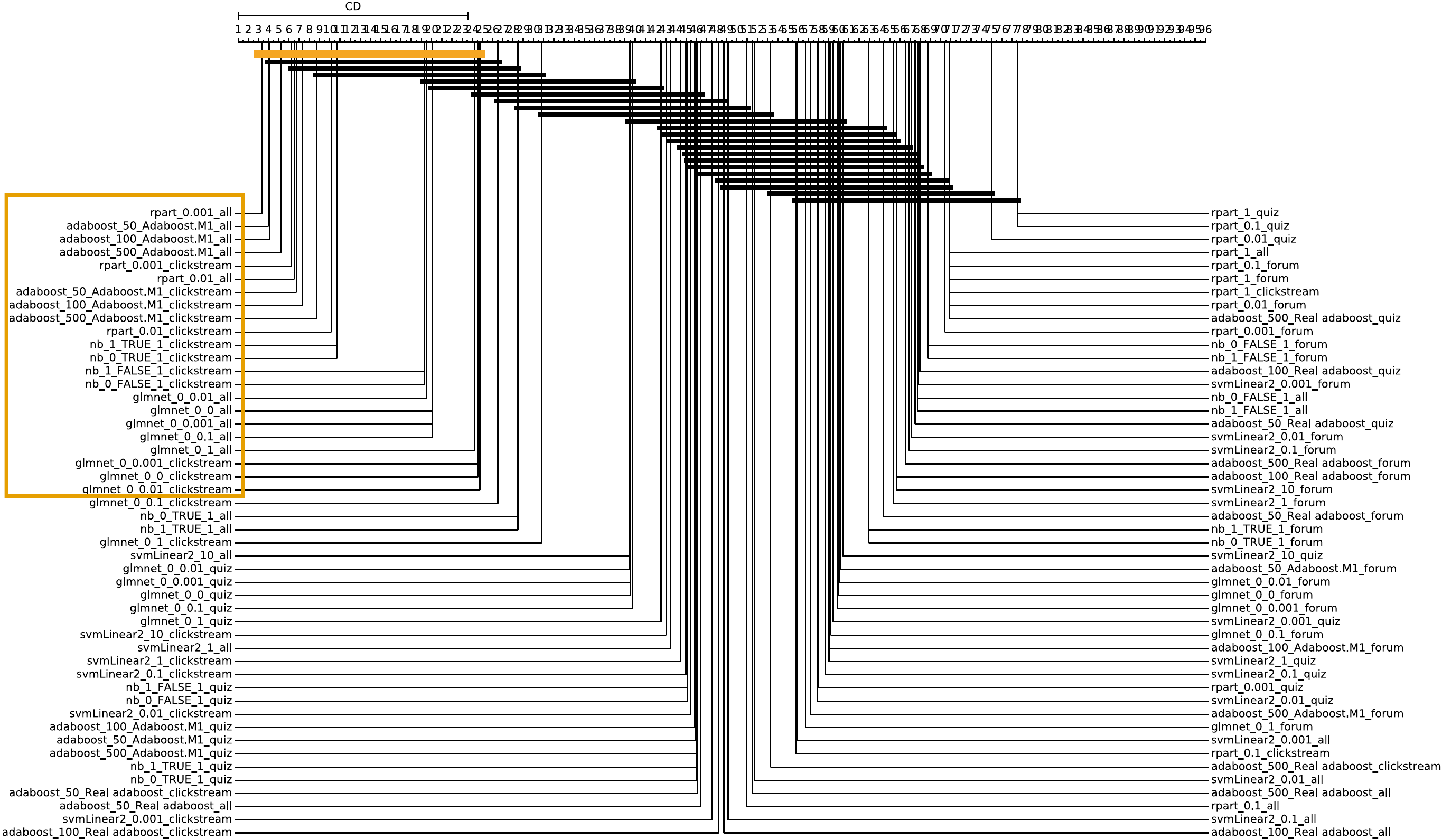}
            \caption{Critical Difference (CD) diagram for models evaluated in this experiment. The bold CD bars connect models statistically indistinguishable at $\alpha = 0.05$; $\widetilde{\mathcal{F}}_{NHST}$ is indicated in orange. This figure illustrates the breakdown of the CD diagram with large numbers of comparisons: the diagram is difficult to read and interpret, with large groups of statistically indistinguishable models.}
            \label{fig:cd-diagram}
        \end{figure}

\begin{figure}[t!]
    \centering
    \includegraphics[width=\textwidth]{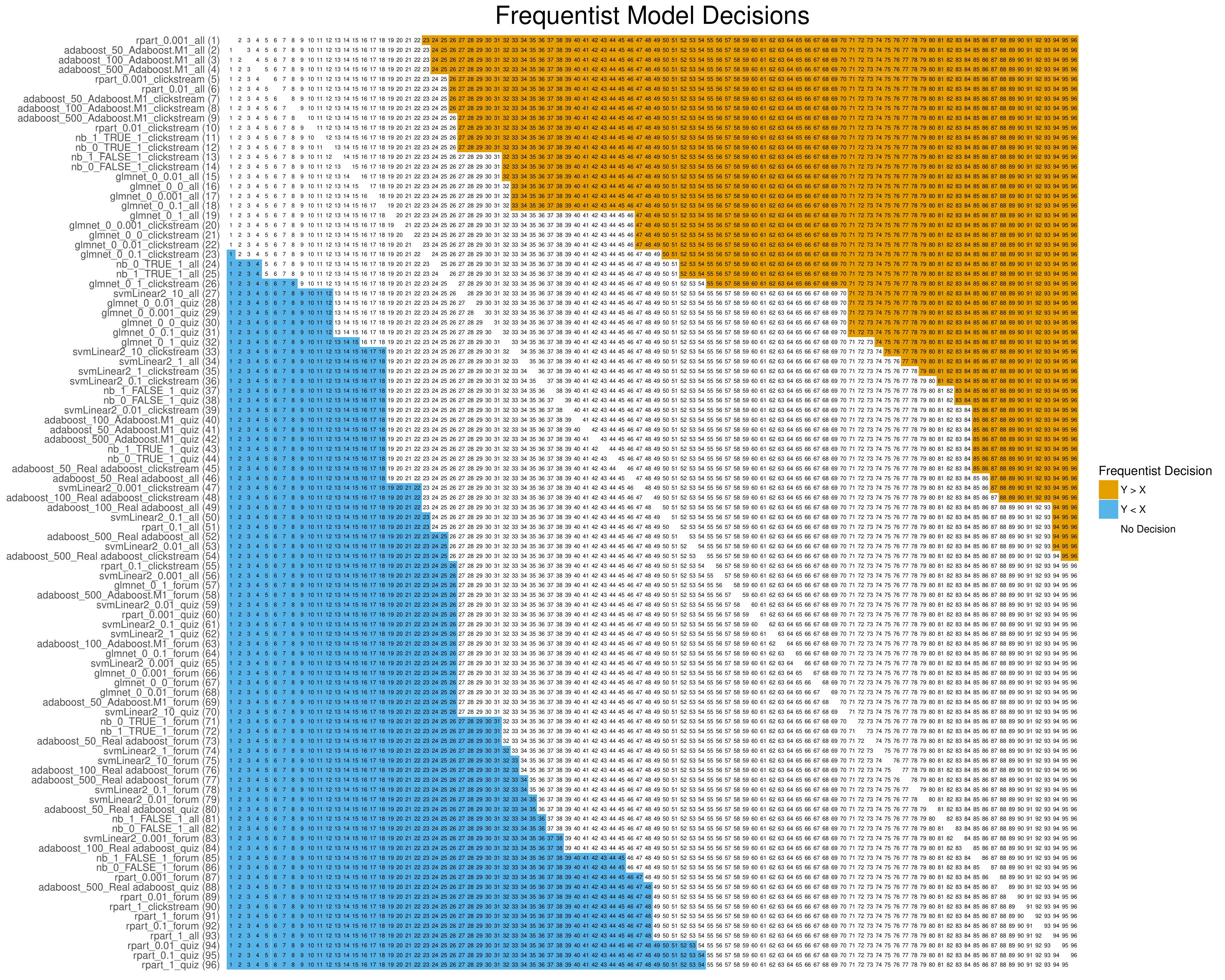}
    \caption{A ``windowpane'' plot showing decisions for frequentist method (adapted from \cite{Mullis1993-tj} via \cite{Gelman2012-ys}). All models are shown along each axis in identical order of decreasing performance. Cells are labeled with the rank of the model along the x axis.}~\label{fig:frequentist-windowpane}
\end{figure}

The NHST procedure recommended in section \ref{sec:nhst} compares models via a two-stage, nonparametric test. Recall that first, the Friedman test (Equation \ref{eqn:friedman}) is applied. If this test indicates a significant difference (as it did in this experiment), a post-hoc Nemenyi test is conducted on all pairwise comparisons to determine where significant differences between individual models may exist. Typically, the results of this procedure are reported using a Critical Difference (CD) diagram, shown in Figure \ref{fig:cd-diagram}. However, the CD diagram is difficult to interpret with a large number of models, which is one way in which this procedure breaks down under large numbers of comparisons; we instead present a ``windowpane plot'' of the results in Figure \ref{fig:frequentist-windowpane}.

The models in $\widetilde{\mathcal{F}}_{NHST}$ are shown in Table \ref{tab:frequentist-best-mods}, and correspond to the set of models in Figure \ref{fig:cd-diagram} linked by an orange CD bar. They are also shown in the windowpane plot in Figure \ref{fig:frequentist-windowpane} as the models which have white-colored cells in the top row (indicating that the models are statistically indistinguishable from the top-ranked model).

\begin{table}[!t]
\centering
\hspace*{-4cm}\begin{tabular}{l l l p{1.1cm} p{1.1cm} p{1.1cm} p{1.1cm}}
\hline 
\textbf{Algorithm} & \textbf{Feature Type} & \textbf{Hyperparameters} & \textbf{Avg. Rank} & \textbf{Avg. AUC} & \textbf{Diff. In Ranks} & \textbf{Diff. In AUC} \\ \hline 
CART & All & $cp = 0.001$    &3.376  & 0.901  & NA& NA \\
Adaboost & All & $NIter = 50$,  Boosting = M1  & 3.978 & 0.899 & -0.602 & 0.002 \\
Adaboost & All & $NIter = 100$,  Boosting = M1  & 4.118 & 0.899 & -0.742 & 0.002 \\
Adaboost & All & $NIter = 500$,  Boosting = M1  & 5.198 & 0.897 & -1.822 & 0.004 \\
Adaboost & Clickstream & $NIter = 50$,  Boosting = M1   & 6.725 & 0.89 & -3.349 & 0.011 \\
Adaboost & Clickstream & $NIter = 100$,  Boosting = M1  & 7.344 & 0.889 & -3.968 & 0.012 \\
Adaboost & Clickstream & $NIter = 500$,  Boosting = M1  & 8.704 & 0.887 & -5.328 & 0.014 \\
Na{\"i}ve Bayes & Clickstream & $fL = 1$, Kernel = True & 10.708 & 0.872 & -7.332 & 0.029 \\
Na{\"i}ve Bayes & Clickstream & $fL = 0$, Kernel = True & 10.708 & 0.872 & -7.332 & 0.029 \\
Na{\"i}ve Bayes & Clickstream &  $fL = 1$,  Kernel = False  & 19.288 & 0.788 & -15.911 & 0.113 \\
Na{\"i}ve Bayes & Clickstream &  $fL = 0$,  Kernel = False  & 19.288 & 0.788 & -15.911 & 0.113 \\
L2LR & All & $\lambda = 0.01$  & 19.521 & 0.78 & -16.145 & 0.121 \\
L2LR & All & $\lambda = 0$ & 20.036 & 0.779 & -16.66 & 0.121 \\
L2LR & All & $\lambda = 0.001$ & 20.036 & 0.779 & -16.66 & 0.121 \\
L2LR & All & $\lambda = 0.1$ & 20.039 & 0.778 & -16.662 & 0.123 \\
L2LR & All & $\lambda = 1$ & 24.258 & 0.752 & -20.882 & 0.149 \\
L2LR & Clickstream & $\lambda = 0.001$  & 24.578 & 0.75 & -21.202 & 0.151 \\
L2LR & Clickstream & $\lambda = 0$ & 24.578 & 0.75 & -21.202 & 0.151 \\
L2LR  & Clickstream & $\lambda = 0.01$ & 24.744 & 0.75 & -21.368 & 0.151 \\ \hline 
\end{tabular}\hspace*{-4cm}
\caption{Family of models $\widetilde{\mathcal{F}}_{NHST}$ which are statistically indistinguishable from the ``best'' model with differences in average rank less than the critical difference of $CD = 22.5936$ relative to the highest-ranked model $\widetilde{f}_{1,NHST}$. All differences are relative to $\widetilde{f}_{1,NHST}$, which has average rank 3.376, average AUC 0.901.}~\label{tab:frequentist-best-mods}
\end{table}

The large size of $\widetilde{\mathcal{F}}_{NHST}$, which consists of 19 models, is due to the large critical difference CD, computed via Equation $\ref{eqn:CD}$. This is the NHST's control for the number of comparisons relative to the number of datasets. The large number of hypotheses in this experiment drastically reduces the NHST's ability to discriminate between closely-ranked models. As Figure \ref{fig:cd-diagram} shows, the orange CD bar connecting the models in $\widetilde{\mathcal{F}}_{NHST}$ occupies nearly 25\% of the length of the number line of possible rankings. The procedure's discrimination could be improved if the procedure used fold-level data (which would increase the sample size $N$), but this would require applying a correction for the overlapping training samples. Such a correction to this test has not, to the authors' knowledge, been theoretically evaluated.

In sections \ref{sec:nhst} and \ref{sec:bayesian-estimation}, we discussed how NHST procedures fail to differentiate between the \textit{magnitude} and the \textit{uncertainty} of the effects under examination during model evaluation. This is reflected in the results shown in Table \ref{tab:frequentist-best-mods}. The NHST procedure does not discriminate between models with large differences in performance (high magnitude) but high variability (high uncertainty), and models with small differences in performance (low magnitude) but also low variability (low uncertainty). This yields an $\widetilde{\mathcal{F}}_{NHST}$ where models may either show (i) large differences in average performance from the best model, but with large enough variability that this might not constitute a real effect, or (ii) models with small average differences but low variability. 

As an example, consider the Adaboost models in Table \ref{tab:frequentist-best-mods} (the six highest-performing models after the decision tree). For these models, the magnitude is quite small, with the difference in AUC from the highest-performing model never greater than 0.01, and we might agree with the NHST's decision to include these in $\widetilde{\mathcal{F}}_{NHST}$. In contrast, each of the logistic regression models shown in the lower portion of Table \ref{tab:frequentist-best-mods} has an average AUC more than 0.1 \textit{worse} than the highest-performing model -- a magnitude large enough to be considered practically important. However, the NHST treats both models as equivalent to $f_{1}$. A user simply presented with $\widetilde{\mathcal{F}}_{NHST}$ would have no way of differentiating between the Adaboost models, which achieved comparable performance to $\widetilde{f}_{1,NHST}$ with low variability, and the logistic regression models, which achieved inferior average performance but higher variability.

Figure \ref{fig:frequentist-windowpane} also shows that the NHST procedure makes decisions in only 44.08\% of the pairwise comparisons conducted (those indicated by nonwhite cells). This provides little information about many pairwise model comparisons; it will make fewer decisions as the number of comparisons $k$ grows for a fixed number of datasets $N$. This illustrates how the frequentist method is only able to make decisions when the observed differences in performance are large (or have low variability) and the hypothesis space moderate, with significant differences only detected between the highest- and lowest-performing models in the sample. This leaves little room for interpretation of the relative difference between, for example, models using quiz vs. forum features (and leads us to conclude that, in most cases, all we can do is draw no conclusion based on the observed data).

Finally, we must correct an abuse of our notation of $\mathcal{F}$. While we would like to conclude that the models shown in Table \ref{tab:frequentist-best-mods} are the ``best'' models, and while this is often implied in the analysis of NHST procedures, to do so would be incorrect. Instead, we \textit{are not able to draw any conclusions regarding the differences in performance between these models} using the NHST procedure. For these models, ``the experimental data is not sufficient to reach any conclusion'' \citep[][pp. 14]{Demsar2006-cx}. This is the inferential equivalent of not being able to reject the null hypothesis, and is why we use the notation of $\widetilde{\mathcal{F}}_{NHST}$ instead of $\mathcal{F}_{NHST}$. A frequentist approach can never \textit{prove} the the null hypothesis of equivalent performance \cite{Corani2017-nx, Kruschke2013-vv}, and instead only produces inconclusive results. NHST can therefore never positively identify $\mathcal{F}_{NHST}$ as defined. 

This is a scientifically unsatisfying result, and it is crucial to the argument of this work: the NHST does not tell us about what we want to know about our experiment. In fact, it allows us only to draw conclusions conditional on a \textit{nil} hypothesis that is almost certainly false: that all of the 96 models evaluated above have \textit{exactly equivalent} performance to all other models. Estimating the probability of observing our results conditional on this hypothesis yields little useful information about the observed data and no information about $\mathcal{F}$, or about pairwise comparisons between more than half of the models in our experiment.

\subsection{Bayesian Model Evaluation Method}

Recall that the Bayesian method, described in Section \ref{sec:bayesian-estimation}, uses the differences in AUC for each pair of models on each cross-validation fold to estimate $\theta = (P(X > Y), P(ROPE), P(X < Y))$. ROPE is the ``region of practical equivalence,'' and denotes that the difference between $X$ and $Y$ is smaller than some pre-specified threshold. This can be thought of as creating a posterior plot, as in Figure \ref{fig:posterior-plot-example}, for each pair of models, and using this posterior distribution to compute each probability in $\theta$. For this experiment, as in previous work using this procedure \cite{Benavoli2017-ff}, we use $ROPE = 0.01$; that is, models are considered practically equivalent if the difference in their AUC is smaller than 0.01.

\begin{table}[]
\centering
\hspace*{-4cm}\begin{tabular}{l l l p{1.1cm} p{1.1cm} p{1.1cm} p{1.1cm}} \hline 
  \textbf{Algorithm} & \textbf{Feature Type} & \textbf{Hyperparameters}  & \textbf{Avg. Rank} & \textbf{Avg. AUC} & \textbf{Diff. In Ranks} & \textbf{Diff. In AUC} \\ \hline 
CART &All & $cp = 0.001$ & 3.376   & 0.901  & NA    & NA      \\  
Adaboost & All & $NIter = 50$,  Boosting = M1 & 3.978 & 0.899 & -0.602 & 0.002 \\
Adaboost & All & $NIter = 100$,  Boosting = M1 & 4.118 & 0.899 & -0.742 & 0.002 \\
Adaboost & All &  $NIter = 500$,  Boosting = M1 & 5.198 & 0.897 & -1.822 & 0.004 \\ \hline                  
\end{tabular}\hspace*{-4cm}
\caption{Family of models $\mathcal{F}_{Bayes}$.}~\label{tab:bayesian-best-mods}
\end{table}

The Bayesian procedure thus allows us to differentiate between cases where the magnitude of the difference is small (the difference $|X - Y| \leq ROPE$, and therefore the models are ``practically equivalent''), and cases where this magnitude is large ($|X - Y| > ROPE$, the model performance is meaningfully different). This is useful because we tend only to care about the latter: if the difference in performance of two models is small, we might consider other aspects of the model, such as training time or interpretability, in order to select which to use.

\begin{figure}[t!]
    \centering
    \includegraphics[width=\textwidth]{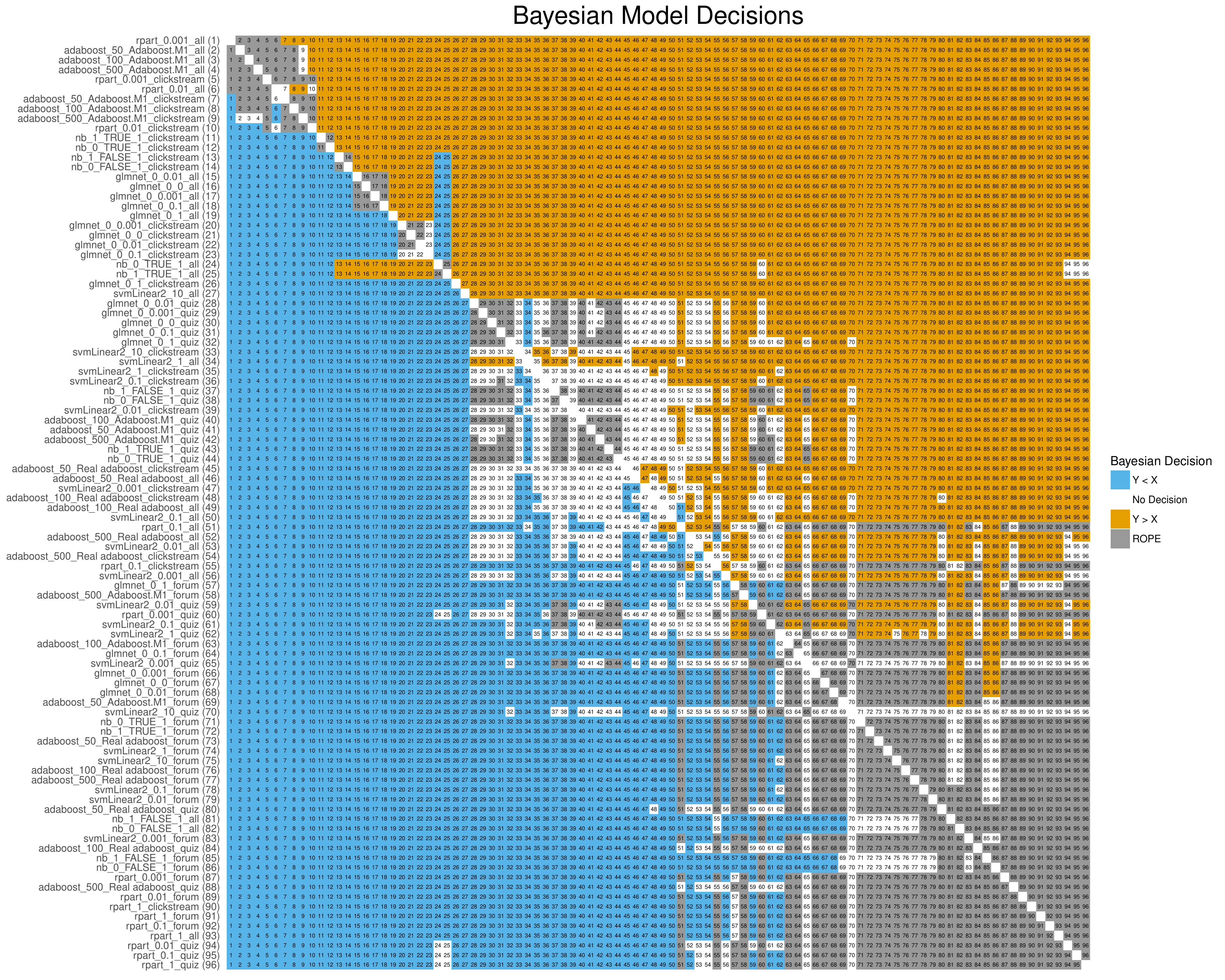}
    \caption{Windowpane plot showing $\mathcal{F}_{Bayes}$. The Bayesian method estimates the posterior probability that two models are within the ``region of practical equivalence'' (ROPE), which is not estimable via NHST.}~\label{fig:bayesian-windowpane}
\end{figure}

We present $\mathcal{F}_{Bayes}$ in Table \ref{tab:bayesian-best-mods} and Figure \ref{fig:bayesian-windowpane}. The Bayesian procedure returns a more precise family of ``best'' models than NHST: $\mathcal{F}_{Bayes}$ contains only four models. All models in $\mathcal{F}_{Bayes}$ have a $P(ROPE)$ greater than the decision threshold of 0.95. This threshold value is selected to replicate \cite{Benavoli2017-ff}; adjustments of the threshold even to 0.999 had minimal effects on the decision in most cases (the model has high confidence, given the data).

Additionally, there is an important epistemic distinction between $\mathcal{F}_{Bayes}$ and its frequentist counterpart $\widetilde{\mathcal{F}}_{NHST}$. Using NHST, recall that we were not able to conclude anything about the models in $\widetilde{\mathcal{F}}_{NHST}$: we cannot reject the null hypothesis of equivalence, but we do not \textit{prove} or provide evidence that these models are equivalent. However, the Bayesian procedure directly estimates the probability of two models being equal: this is $P(ROPE)$, the probability that models $X$ and $Y$ are practically equivalent. Being able to model practical equivalence is quite beneficial for a large hypothesis space: there are many comparisons for which the model performance is reliably within the region of practical equivalence (ROPE), and the Bayesian approach can assign high probability to ROPE in such cases. In contrast, the frequentist paradigm can never conclude that no difference exists; indeed, given infinite data, as noted above, the frequentist paradigm will \textit{always} conclude that a significant different exists (even when this difference is very small).

$\mathcal{F}_{Bayes}$ is smaller than $\widetilde{\mathcal{F}}_{NHST}$, in part because $\mathcal{F}_{Bayes}$ is sensitive to the magnitude of the difference between models. Only those models with small differences in performance relative to the best model $f_1$ are in $\mathcal{F}_{Bayes}$; those with large, but highly variable, differences from $f_1$ are not). The Bayesian method allows us to make substantive conclusions about those models. These conclusions are not conditional on the assumption of a null hypothesis of equivalent performance (which is almost certainly incorrect in most cases), and is instead only conditional on the data we observe.

While the na{\"i}ve average approach may be favored for its ability to yield a small, precise $\mathcal{F}$ consisting of only a single model, the Bayesian approach nearly matches the precision of using a na{\"i}ve average approach in this case. However, the Bayesian method does so without making unwarranted assumptions about the significance of observed differences -- recall that the na{\"i}ve average method simply assumes that \textit{all} differences are significant. Instead, the Bayesian approach uses hierarchical modeling to directly estimate the probability that each pair of models is practically equivalent, given their observed performance. 

Above, we demonstrated how the NHST method is especially sensitive to the number of comparisons, for a fixed number of datasets, in order to control the size of the cricial difference. Additionally, this requires researchers to track, report, and account for the full scope of comparisons performed in the course of an experiment -- despite the fact that increasing the number of comparisons will reduce their ability to detect a statistically significant effect (by increasing the size of the CD). For Bayesian model evaluation, large numbers of comparisons are generally not considered to be a concern \cite{Gelman2012-ys}, because the concept of a Type I error does not exist under a Bayesian framework. Additionally, the hierarchical model accounts for the variability within and across models, reducing the estimated effect sizes (in this case, differences in observed performance) as necessary \cite{Gelman2012-ys}.

\subsection{Implications for Predictive Modeling in MOOCs}

In previous sections, we focused on the methodological implications of our case study. In this section, we evaluate the practical conclusions from the case study. We intended this experiment to be useful, interesting, and realistic, and we present the main findings with respect to predictive modeling in MOOCs here (using the results of the Bayesian analysis).

Table \ref{tab:bayesian-best-mods} and Figure \ref{fig:bayesian-windowpane} both show that all models in $\mathcal{F}_{Bayes}$ use \textit{all} features -- clickstream, forum, and assignments: models that used these combined features made better predictions than those with only one of the feature sets. Additionally, Figure \ref{fig:bayesian-windowpane} provides further information on which individual feature sets (and data sources) demonstrated the best performance: by inspecting the model rankings, and the large ``blocks'' of models which are practically equivalent, we see that models using clickstream features consistently outperformed models using forum or assignment features, and in many cases clickstream-only models were competitive with, or practically equivalent to, models using all features. In most cases, models with forum or assignment features produced practically equivalent performance, regardless of algorithm and hyperparameters used, as indicated by the large ``ROPE'' block in the lower right-hand section of Figure \ref{fig:bayesian-windowpane}. This suggests that no model was able to extract substantially more information than any other from the forum and assignment features; we hypothesize that this is because these features are highly sparse and are only available for a small subset of students who choose to complete those optional activities.

In terms of particular algorithms, $\mathcal{F}_{Bayes}$ shows that nonparametric tree-based models show the best performance across the large and diverse sample of MOOCs used here. This comports with current practice in learning analytics, where tree-based methods are already widely used \cite{Gardner2018-lp}.

Additionally, Figure \ref{fig:bayesian-windowpane} demonstrates that, for many algorithms, hyperparameter tuning appears to have had little effect, especially relative to the obvious effects of feature type: many different tunings of otherwise-identical models are practically equivalent, or even adjacent in the average model rankings. This suggests that future modeling efforts might achieve the most substantial performance improvements from feature engineering using a rich data source, such as the clickstream, and not from extensively tuning sophisticated algorithms on sparse or uninformative feature sets. Using a reasonable default hyperparameter setting might, in most cases, be sufficient to realize most of the models' performance benefits if the feature set is of high quality, as our experiment included the default hyperparameters for each model.

Often, predictive models in learning analytics require \textit{interpretability} in order to communicate the findings to stakeholders or inspect the model itself. This case study demonstrates several findings of interest here. First, it shows that a highly interpretable model (CART) can achieve excellent performance. Other interpretable models, such as the L2LR model (models 15-23 in Figure \ref{fig:bayesian-windowpane}) also achieved relatively strong performance. Second, the procedure demonstrated here could be implemented with any class of candidate models; if a researcher wished only to use a certain type of highly interpretable algorothm (such as the logistic regression models considered), these could be used as the candidate model space instead. Third, in a case where interpretability is perhaps secondary to predictive performance but still important, the most interpretable model could be selected from $\mathcal{F}$; in the case of $\mathcal{F}_{Bayes}$, this would likely be the CART model.

\section{Conclusion and Notes for Practice}\label{sec:conclusion}

In this work, we are concerned with advancing the state of the learning analytics field with respect to the evaluation of predictive models. We presented the results of a comprehensive literature review to assess the state of the practice in the field, and presented an overview of several techniques for model evaluation. By applying these results to a realistic and comprehensive case study using a large set of MOOC data, we demonstrated the differences in substantive conclusions supported by each method. In particular, we demonstrated the power of Bayesian model evaluation to draw highly precise, informative conclusions about the performance of feature-algorithm-hyperparamter combinations, even across a large space of candidate models. Under the Bayesian model evaluation method, our case study specifically demonstrated the importance of feature extraction to model performance, and in particular demonstrated the predictive performance of clickstream-based features for dropout prediction (and the relatively poor performance achieved by using only forum- or assignment-based features, regardless of the statistical model used).

We hope that this work contributes to a growing movement in the field toward addressing several of the existing barriers to rigorous model evaluation in learning analytics and across domains, including those described in Section \ref{sec:why-model-eval-rare}.

While the fields of learning analytics and educational data mining are only a decade old\footnote{Based on annual conferences in the area, though we note that educational data mining itself has roots in the much older Artificial Intelligence in Education (AIED) community.} the techniques we have presented here borrow from the work of others done over the last 15 years in the broader machine learning community, and largely reflect the evolution of the field of machine learning as a whole as it develops methods to address new challenges brought by the expansion of data science and computing. 

With the growth in the size of datasets available (e.g. MOOCs), and the ability to run thousands of permutations of analyses on desktop hardware, we are concerned with the lack of rigor when selecting and reporting on the ``best'' predictive model. This is especially important as there a number of pragmatic elements when operationalizing predictive models -- computational speed, robustness in the face of missing data, and interpretability -- all of which might influence adoption ability. In this work we have shed light on techniques which can help inform these choices, and we look forward to the growth of rigor in the community as a result. 

\bibliographystyle{abbrv}
\bibliography{MSI_0_2}

\end{document}